\UseRawInputEncoding
\documentclass[aps,prc,twocolumn]{revtex4-1}

\usepackage{hyperref,amsmath,graphicx,url,xcolor}

\begin{document}

\title{Resonance production in PbPb collisions at 5.02 TeV via hydrodynamics and hadronic afterburner}

\author{D. Oliinychenko$^1$}
\author{C. Shen$^{2,3}$}

\address{1 
Institute for Nuclear Theory, University of Washington, Seattle, WA, 98195, USA}
\address{2 Department of Physics and Astronomy, Wayne State University, Detroit, Michigan 48201, USA}
\address{3 RIKEN BNL Research Center, Brookhaven National Laboratory, Upton, NY 11973, USA}

\begin{abstract}
Using a relativistic hydrodynamics + hadronic afterburner simulation we explore resonance production in PbPb collisions at 5.02 TeV, and demonstrate that many resonance yields, mean transverse momenta, and flows are very sensitive to the late stage hadronic rescattering. Out of all measured resonances $\Lambda(1520)$ is affected strongest by the hadronic rescattering stage, which allows to estimate its duration, and even constrain branching ratios of $\Sigma^* \to \Lambda(1520)\pi$ decays. Strong suppression of $\Lambda(1520)$, which in vacuum has a lifetime of 12.6 fm/$c$, is explained by its small lifetime in a hadronic medium, between 1 and 2 fm/$c$  at temperatures between 100 and 150 MeV. We find that some resonances like $\Delta(1232)$, $f_0(980)$, $a_0(980)$, $\Lambda(1405)$ are enhanced rather than suppressed by the afterburner. 
\end{abstract}

\maketitle

\section{Introduction}
Resonance production in ultra-relativistic heavy-ion collisions is rather well-studied experimentally. From measurements of resonances by STAR collaboration in pp and AuAu collisions at 200 GeV \cite{Adams:2004ep,Adams:2006yu} and ALICE collaboration in pp, pPb, and PbPb at 2.76 and 5.02 TeV \cite{Acharya:2018qnp,Abelev:2014uua,ALICE:2018ewo,Adamova:2017elh,ALICE:2018ewo} it is known that midrapidity yield ratios such as $\rho^0/\pi$, $K^{*0}/K^-$, $\Lambda(1520)/\Lambda$ are suppressed in larger colliding systems compared to smaller ones. Some ratios like $\varphi/K$ or $\Sigma(1385)/\Lambda$ remain approximately the same in central and peripheral collisions. It is well established that resonance production is very sensitive to the late stage of the fireball expansion, where hadronic rescattering occurs (see e.g. \cite{ALICE:2018ewo}). Models without hadronic rescattering fail to reproduce resonance suppression in central collisions, while hydrodynamics + transport simulations reproduce it rather well \cite{Knospe:2015nva}. However qualitative understanding of the phenomenon is missing: it is challenging to even predict if a resonance $X$ is suppressed more or less than a resonance $Y$ without running a numerically expensive simulation. A popular idea to use a vacuum lifetime of the resonance as a predictor fails: $\Lambda(1520)$ with lifetime of $12.6 \pm 0.8$ fm/$c$ is suppressed rather strongly, $\varphi$ with lifetime of $46.4 \pm 0.14$ fm/$c$ is not suppressed, while $K^{*0}(892)$ with the lifetime of $4.17 \pm 0.04$ fm/$c$ is moderately suppressed. Similarly, using a regeneration cross section as a predictor fails \cite{Adams:2006yu}. More promising are two opposite ideas: a ballistic one based on Knudsen number (ratio of resonance mean free path to fireball size) and a statistical one based on the concept of partial chemical equilibrium \cite{Hirano:2002ds, Shen:2010uy, Motornenko:2019jha}.

Let us elaborate, why both could potentially be useful predictors. During part of the evolution, the fireball is locally equilibrated, both chemically and kinetically. Therefore hydrodynamics can be applied to describe it. Due to the expansion, the density drops, and one can start viewing a fireball as a system with multiple ongoing hadronic reactions. After a certain moment, called chemical freeze-out, the yields of stable hadrons (with respect to the strong interaction, e.g., $\pi$, $K$, $p$, $\Lambda$, $\Xi$, $\Omega$) are not changed substantially, because stable hadron number-changing reactions such as
$NN \leftrightarrow N\Delta \leftrightarrow NN \pi$, $\Lambda K \leftrightarrow \Xi \pi$, $p\bar{p} \leftrightarrow 5 \pi$ either exhibit small rates, or their forward and reverse rates are close to equal. Resonance formations and decays such as $N\pi \to \Delta \to N\pi$ do not change the yields of stable hadrons, while reactions like $\pi\pi \to f_2 \to K^+K^-$ seem to be either relatively rare or having similar rates in forward and reverse directions. Sharp chemical freeze-out simultaneous for all hadron species is an idealization, but it has proven to be a useful concept both qualitatively and quantitatively. Statistical Hadron Resonance Gas model based on this concept describes stable hadron yields rather accurately; see \cite{Andronic:2017pug} for details. The reactions that change resonance yields can proceed after chemical freeze-out. Resonances can collide and be excited to higher mass resonances (which can also be understood as smaller in-medium width than vacuum width), or their decay products can re-scatter in the hadronic medium. These processes are usually simulated by a non-equilibrium hadronic transport.

In the limit of small cross-sections with other hadrons, a resonance will either escape the fireball or collide and possibly disappear. The probability of disappearing is proportional to the Knudsen number $Kn = \lambda/L$, where $\lambda$ is the mean free path of the resonance and $L$ is system size. This argument would explain the experimentally observed suppression of resonances in central Pb+Pb or Au+Au collisions compared to peripheral ones -- larger system size means smaller escape chance. This consideration has two drawbacks: (i) it does not take regeneration of resonances into account, and (ii) its initial assumption of $Kn > 1$ is only fulfilled for few resonances, such as $\varphi$. An alternative idea to Knudsen number consideration is that reactions of resonance formation occur at high enough rates to keep resonances in relative thermodynamic equilibrium with stable hadrons.  Such an idea was developed in \cite{Hirano:2002ds, Motornenko:2019jha}, and explains measured resonances rather well, except $\Lambda(1520)$ suppression is somewhat underestimated. The centrality dependence of the resonance suppression in this model originates from different temperatures of kinetic freeze-out for different centralities. The main drawback of such a model is an assumption of rapid kinetic freeze-out simultaneous for all species, but it is a necessary sacrifice for model simplicity. How well this model agrees with full hydrodynamics + afterburner simulation is a question we would like to explore.

The goals of this work are threefold. First, we would like to test the theoretical ideas above using hydrodynamics + hadronic afterburner simulation. Second, new data on resonance production from ALICE collaboration at 5.02 TeV are expected soon, and we provide theoretical predictions for resonance yields, mean transverse momenta, and elliptic flows. Third, we explore what one can learn about the fireball and resonances themselves from these measurements.  The work is organized as follows: in Sec. \ref{sec:Methodology} we explain our hydrodynamics + afterburner simulation, in Section \ref{sec:Results} we
\begin{itemize}
    \item discuss the obtained yields of stable hadrons and resonances, their mean transverse momenta, and elliptic flows as a function of centrality
    \item explore what one can learn from $\Lambda(1520)$ measurement: about hadronic stage duration and currently unknown branching ratios of  $\Sigma^* \to \Lambda(1520)\pi$ decays
    \item investigate what quantities can be good predictors of the resonance suppression
    \item consider resonances that are enhanced rather than suppressed in our simulation
\end{itemize}
and finally, in Section \ref{sec:Summary} we briefly summarize our findings.

\section{Methodology} \label{sec:Methodology}

\begin{figure}
    \centering
    \includegraphics[width=0.95\linewidth]{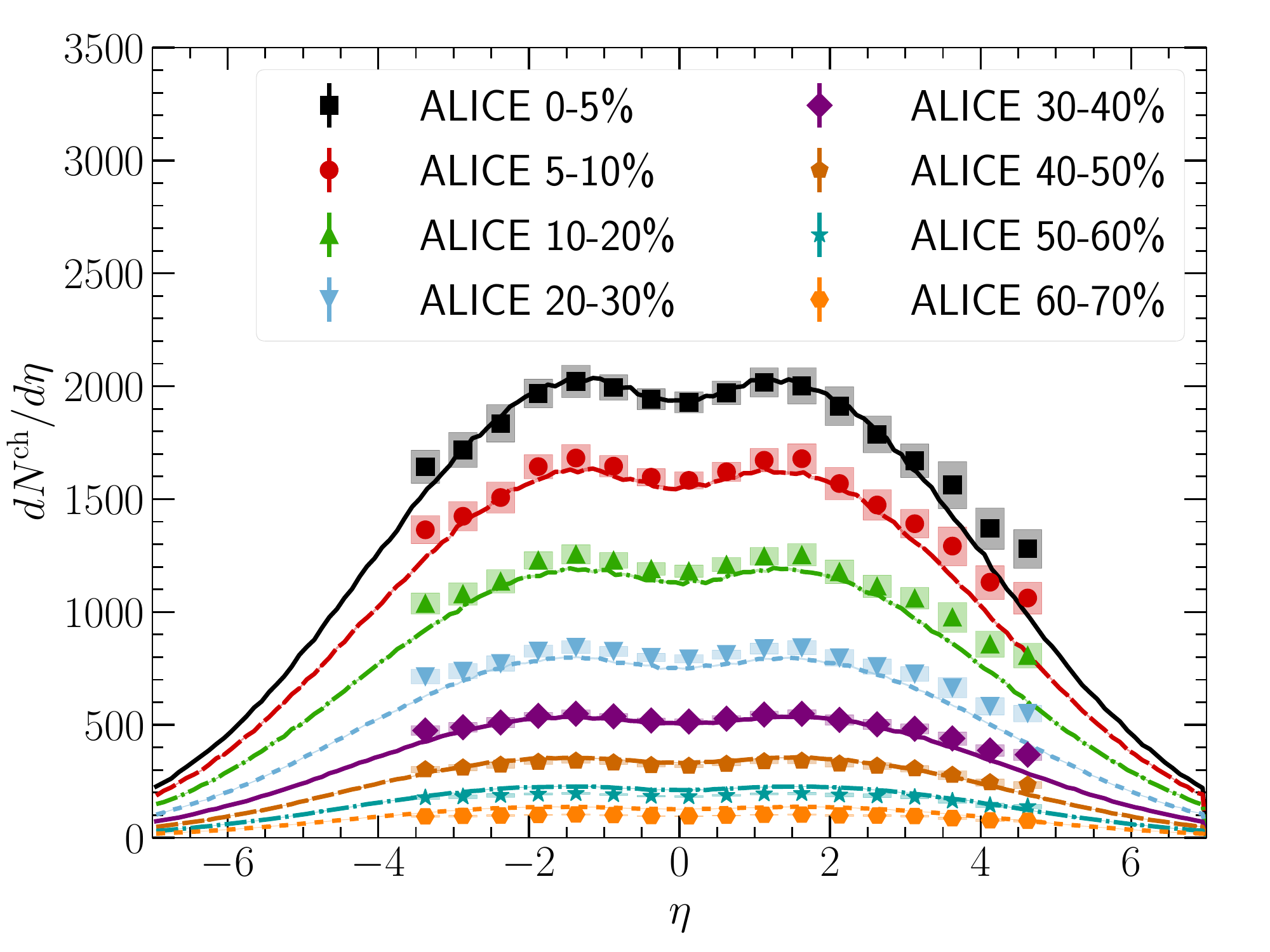}
    \caption{(Color online) Charged hadron multiplicity distribution from our 3D hydrodynamics + hadronic transport model compared with the ALICE measurements \cite{Adam:2016ddh} in different centrality bins.}
    \label{fig:dNdeta}
\end{figure}

\begin{figure*}
    \centering
    \includegraphics[width=0.45\textwidth]{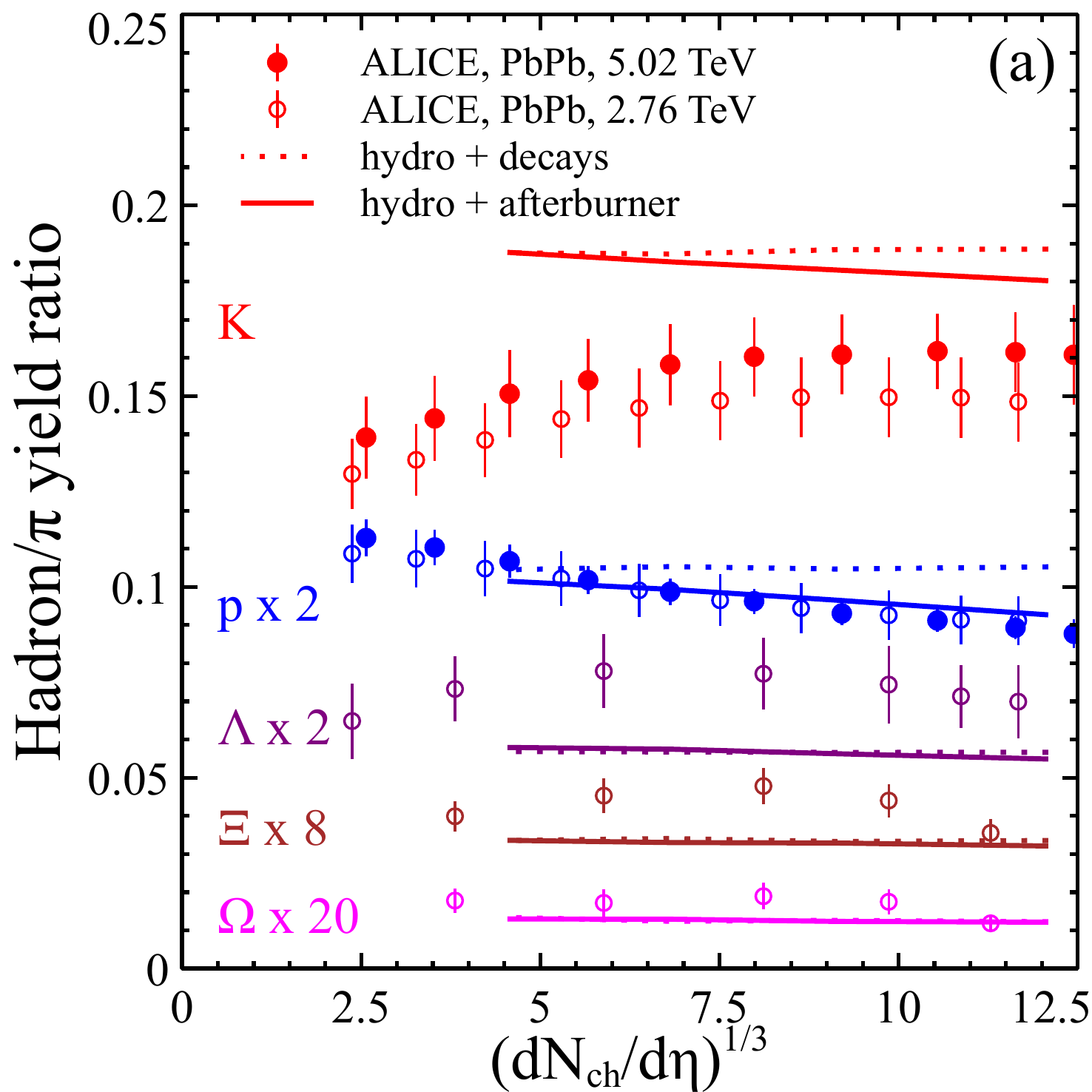}
    \includegraphics[width=0.45\textwidth]{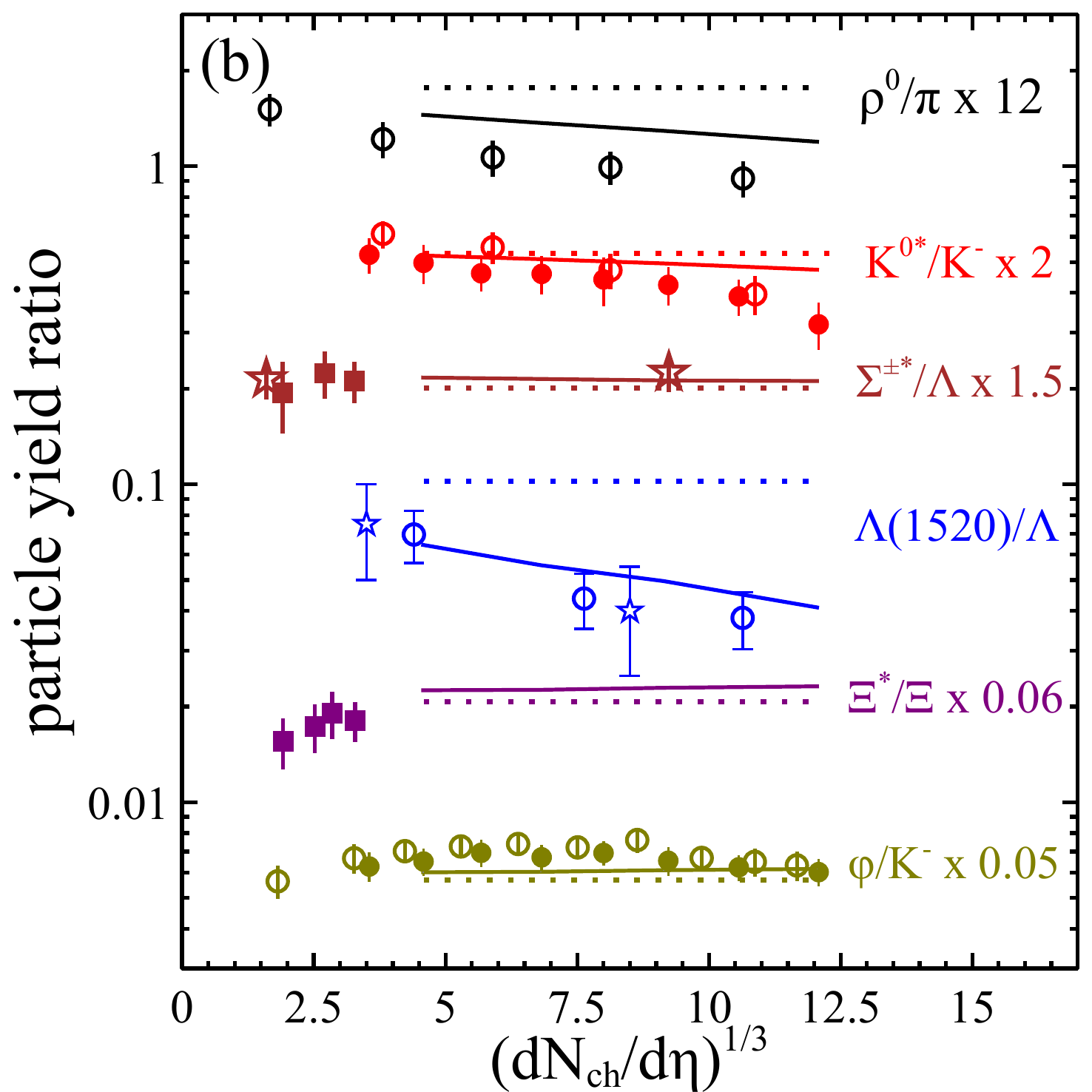}
    \caption{Midrapidity yield ratios of stable hadrons (a) and selected resonances (b) in PbPb collisions at 5.02 TeV. Ratios are shown as a function of a charged particle multiplicity  per unit of pseudorapidity, $dN_{ch}/d\eta$, at $\eta = 0$. Dotted lines correspond to hydrodynamics simulation and resonance decays, full lines stand for the simulation with account of hadronic rescattering after hydrodynamical stage. The results of simulations are compared to experimental data from PbPb collisions at 5.02 TeV (full circles) \cite{Acharya:2019yoi,Acharya:2019qge}, PbPb collisions at 2.76 TeV (open circles) \cite{Abelev:2013vea,Abelev:2013xaa,ABELEV:2013zaa,Acharya:2018qnp,Abelev:2014uua,ALICE:2018ewo}, pPb collisions at 5.02 TeV (squares) \cite{Adamova:2017elh}, and AuAu collisions at 200 GeV (stars) \cite{Adams:2006yu}.}
    \label{fig:yields}
\end{figure*}

We simulate Pb+Pb collisions at 5.02 TeV using a state-of-the-art hybrid (hydrodynamics + hadronic afterburner) approach. Hydrodynamic simulation starts with generating a 3-dimensional initial condition at time $\tau_0 = 1$ fm/$c$. The  energy density and net-baryon density as a function of space and time, $\epsilon(\tau, x, y, \eta_s)$ and $n_B(\tau, x, y, \eta_s)$, are initialized according to parametrizations described in Ref.~\cite{Shen:2020jwv},
\begin{eqnarray}
    && \epsilon (\tau_0, x, y, \eta_s)  = \mathcal{N}_e(x, y) \nonumber \\
    && \!\!\!\!\!\quad \times \exp\left[- \frac{(\vert \eta_s - y_\mathrm{CM}\vert  - \eta_0)^2}{2\sigma_\eta^2} \theta(\vert \eta_s - y_\mathrm{CM} \vert - \eta_0)\right]
    \label{eq:eprof}
\end{eqnarray}
and
\begin{equation}
    n_B(\tau_0, x, y, \eta_s) = f^{+}_{n_B}(\eta_s) T_A(x, y) + f^{-}_{n_B}(\eta_s) T_B(x, y),
    \label{eq:nBprof}
\end{equation}
where longitudinal profiles $f^{\pm}_{n_B}(\eta_s)$ are normalized asymmetric Gaussian,
\begin{eqnarray}
    &&f^{\pm}_{n_B} (\eta_s) = \mathcal{N}_{n_B} \nonumber \\
    && \times \exp\left[ - \frac{(\eta_s \mp \eta_{B,0})^2}{2\sigma_{B, \pm}^2 \theta(\eta_s \mp \eta_{B,0}) + 2\sigma_{B, \mp}^2 \theta(\pm \eta_{B,0} - \eta_s)} \right].
    \label{eq:nBprof_eta}
\end{eqnarray}
The normalization factor $\mathcal{N}_e(x, y)$ and $y_\mathrm{CM}(x, y)$ are functions of the nuclear thickness functions $T_{A,B}(x, y)$ \cite{Shen:2020jwv}. For Pb+Pb collisions at 5.02 TeV, we choose the initial-state parameters to fit the measured charged particle pseudo-rapidity distribution as shown in Fig.~\ref{fig:dNdeta}. The plateau width of the energy density is set as
\begin{equation}
    \eta_0(N_\mathrm{part}) = 2.15 - \left( \frac{N_\mathrm{part}}{416} - 0.5 \right),
\end{equation}
where $N_\mathrm{part}$ varies as a function of centrality.
The other parameters are listed in Table~\ref{table1}.
\begin{table}[ht!]
    \centering
    \begin{tabular}{|c|c|c|c|c|}
    \hline
        $\sqrt{s_\mathrm{NN}}$ (GeV) & $\sigma_\eta$ & $\eta_{B,0}$ & $\sigma_{B,-}$ & $\sigma_{B,+}$ \\ \hline
        PbPb @ 5020  &  2.15 &   6     &   2.0     &   0.1 \\ \hline
    \end{tabular}
    \caption{The model parameters for longitudinal envelope profiles for system's local energy density and net baryon density.}
    \label{table1}
\end{table}

Our simulations use event-averaged smooth initial-state profiles, which do not include event-by-event fluctuations. Using a smooth initial state substantially decreases simulation runtime, and it is justified because we do not consider high order anisotropic flow beyond $v_2$ nor flow fluctuations in this work. The initial energy-momentum tensor is assumed to have a diagonal ideal-fluid form $T^{\mu\nu} = (\epsilon + p)u^{\mu}u^{\nu} - p g^{\mu\nu}$.
At $\tau = \tau_0$, Bjorken flow is assumed: $u^\mu = (\cosh \eta_s, 0, 0, \sinh \eta_s)$. An open-source 3-dimensional relativistic hydrodynamic code \texttt{MUSIC v3.0} \cite{Schenke:2010nt, Schenke:2011bn, Paquet:2015lta, Denicol:2018wdp, MUSIC} is employed to propagate the energy-momentum tensor as a function of $\tau$ and space until energy density for all cells is below $\epsilon_p = 0.2$ GeV/fm$^3$. The equation of state combined with hydrodynamic equations is a lattice QCD based ``NEOS-BSQ'' equation of state $p=p(\epsilon, n_B)$ described in Ref.~\cite{Monnai:2019hkn}. Shear viscous corrections are included with a specific shear viscosity $\eta T/(e + P) = 0.1$, while bulk viscous corrections and baryon number diffusion are neglected. Particlization is performed at a constant energy-density hypersurface, $\epsilon(\tau,x,y,\eta_s) = 0.2$ GeV/fm$^3$. At the collision energy 5.02 TeV at midrapidity the net-baryon density $n_B \approx 0$, and $\epsilon(\tau,x,y,\eta_s) = 0.2$ GeV/fm$^3$ corresponds to the ideal hadron resonance gas temperature $T_{p} \approx 145$ MeV.

The particlization temperature of 145 MeV is somewhat lower than the temperature $156.5 \pm 1.5$ MeV obtained from the hadron resonance gas fits of the stable hadron midrapidity yields \cite{Andronic:2017pug}. At least part of the difference may be connected to the fact that we assign less importance to fitting the multi-strange hadron yields, which would indeed be described better by a higher temperature. This tension between protons and multi-strange baryons is well-known in the hadron resonance gas model (see \cite{Andronic:2018qqt} for a recent attempt to resolve it). The yields of stable hadrons at different centralities are shown in Fig.~\ref{fig:yields}, one can see that our model fits protons rather well, slightly overestimates kaon yield, while $\Lambda$, $\Xi$, and $\Omega$ yields tend to be underestimated. Note that the $\Lambda$ yields in both panels of Fig.~\ref{fig:yields} are in fact $\Lambda + \Sigma^0$ yields both for ALICE and in our simulation. In heavy ion collision experiments a $\Lambda$ originating from $\Sigma^0$ is indistinguishable from primordial $\Lambda$ because of the fast electromagnetic decay $\Sigma^0 \to \Lambda \gamma$, where the lifetime of $\Sigma^0$ is $\tau \approx 14.3 \cdot 10^{-20}$ s~\cite{Patrignani:2016xqp}. In Fig.~\ref{fig:yields}
we show yield ratios to pion yield, because pions are described well by construction: the initial state rapidity profiles are tuned to describe the charged particle pseudo-rapidity distributions, and most of the charged particles at this energy are pions.

In this work we explore two methods of particlization: with account of resonance spectral functions and without. Without spectral functions a usual Cooper-Frye formula is used to compute the spectra from a piece of hypersurface with normal 4-vector $d\sigma_{\mu}$:
\begin{align} \label{Eq:1}
    p^0 \frac{d^3N}{dp^3} = p^{\mu} d\sigma_{\mu} (\mathit{f}_\mathrm{eq}(p, m_0) + \delta \mathit{f}(p, m_0))
\end{align}
Here $N$ is a number of hadrons with momentum $p$, $\mathit{f}_\mathrm{eq}$ is an equilibrium distribution function and $ \delta \mathit{f}$ is a shear-viscous correction, which changes the spectra but does not contribute to yields, because $\int d^3p \frac{p^{\mu} d\sigma^{\mu}}{p^0} \delta \mathit{f}(p, m_0) = 0$ by construction. Here, in Eq.~(\ref{Eq:1}), we underline that the distribution function depends on the pole mass of a hadron $m_0$, but not on its spectral function. Sampling of the particles according to the Eq.~(\ref{Eq:1}) is performed by the the \texttt{iSS} sampler v1.0, which was described and tested in \cite{Shen:2014vra} and is available publicly at \cite{ISS}. In the process of investigation we realized that the account of resonance spectral functions at particlization could potentially change our results. Therefore, we also study particlization with spectral functions:
\begin{align} \label{Eq:2}
    p^0 \frac{d^3N}{dp^3} = p^{\mu} d\sigma_{\mu} (\mathit{f}_\mathrm{eq}(p, m^2) + \delta \mathit{f}(p, m^2)) \mathcal{A}(m^2) dm^2 \,,
\end{align}
where $2 m \mathcal{A}(m^2)$ is a spectral function of a resonance. The spectral functions are taken directly from the \texttt{SMASH} hadronic transport code \cite{Weil:2016zrk}, version 2.0, which we subsequently employ as an afterburner. Therefore, for particlization with spectral functions there is a consistency between the hadron sampler and afterburner. For sampling with spectral functions we utilize a code recently developed at Michigan State University, which we further call \texttt{MSU} sampler \cite{MSU}.

Final-stage hadronic rescatterings and resonance decays are simulated by the \texttt{SMASH} hadronic transport code, which includes elastic collisions, resonance formation and decays, $2 \to 2$ inelastic reactions such as $NN \to N\Delta$, $NN \to N N^*$, $NN \to N\Delta^*$ ($N^*$ and $\Delta^*$ denote all nucleon- and delta-resonances), and strangeness exchange reactions. String formation and its multi-particle decay are also included, but their role is negligible in the case of an afterburner. The \texttt{SMASH} resonance list comprises most of the hadron resonances listed in the Particle Data Group collection \cite{Patrignani:2016xqp} with pole mass below 2.6 GeV. We utilize the public version 2.0 of the \texttt{SMASH} code without any modifications, except when specifically mentioned in the text, for example, when we vary $\Sigma^* \to \pi \Lambda(1520)$ branching ratios.

Unlike in experiments, we do not need to reconstruct resonances by invariant mass distribution or secondary vertex geometry. In our simulations the entire collision history is recorded. We consider resonances as measurable if they did not collide inelastically, and their decay products have reached final state time 100 fm/$c$ without any rescattering, elastic or inelastic, at any point of the decay chain. We check that our results do not change if the end time is increased to 1000 fm/$c$.

\section{Results and discussion} \label{sec:Results}

\subsection{Effects of hadronic rescattering: yields, mean transverse momentum, and flow}
\begin{figure*}
    \centering
    \includegraphics[width=0.7\textwidth]{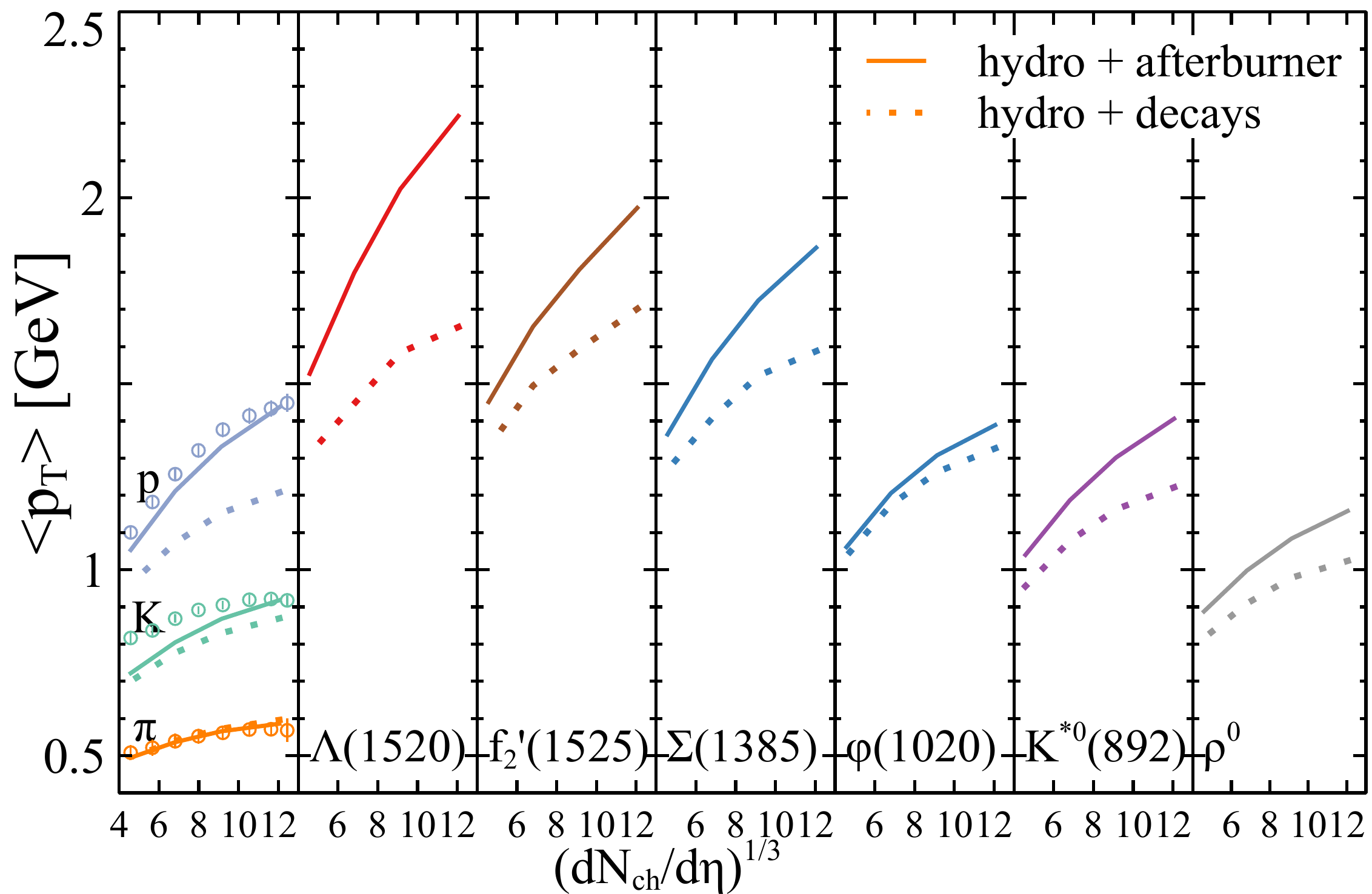}
    \caption{Predictions for mean transverse momentum of resonances in Pb+Pb collisions at 5.02 TeV. Collision centrality is varied, most central events correspond to the largest pseudorapidity density of charged particles $dN/d\eta$ at $\eta = 0$. The difference between hydrodynamics and hadronic afterburner simulations (solid lines), compared to hydrodynamics followed by resonance decays (dashed lines) shows a big role of hadronic afterburner for simulating resonance production. ALICE data \cite{Acharya:2019yoi} are shown with circles. }
    \label{fig:mpt}
\end{figure*}

The first question we study here is the role of the hadronic rescattering stage for the production of stable hadrons and resonances. In Fig.~\ref{fig:yields} one can see that if only resonance decays are performed as a final stage (no rescattering), then the midrapidity yield ratios $K/\pi$, $p/\pi$, $\Lambda/\pi$, $\Xi/\pi$, $\Omega/\pi$ are independent on centrality. This is expected because the ratios depend only on particlization temperature, which we do not change with centrality. In the case of a hadronic afterburner in the last stage, one can see in Fig.~\ref{fig:yields} that proton and kaon yields tend to be slightly suppressed at central collisions, while multi-strange baryons are almost unaffected. Protons are suppressed due to baryon-antibaryon annihilation reactions, $B\bar{B} \to $ pions. In \texttt{SMASH} these annihilation reactions are only implemented in one direction, i.e., multiple pions cannot form a baryon-antibaryon pair. Therefore, the difference of ``hydro + decays'' and ``hydro+afterburner'' for protons in Fig.~\ref{fig:yields} represents an estimate from above for the annihilation effects. One can also see in Fig.~\ref{fig:yields} that the kaon midrapidity yield is affected by the hadronic rescattering. In the model, the trend against centrality is the same as for protons -- slightly smaller kaon yield in the more central events, which may be both due to strangeness exchange reactions, as well as reactions like $KK \to f^* \to \pi\pi$, where $f^*$ denotes a family of mesonic resonances. Although the kaon yields in the model agree with the experiment, the trend we obtain is the opposite. Experimentally $K/\pi$, as well as $\Lambda/\pi$, $\Xi/\pi$, $\Omega/\pi$, are smaller in collisions of smaller systems, which is sometimes referred to as ``strangeness enhancement'' in PbPb relative to pp. One state-of-the-art explanation of strangeness enhancement is that in smaller systems, only a fraction of a fireball (``core'') can be treated hydrodynamically, while the other fraction (``corona'') should be treated ballistically \cite{Kanakubo:2019ogh}. Simulating core-corona separation requires a dynamic initial state, which we do not include in our simulation -- in the core-corona terms, our whole system is assumed to be a core. The system we study is large enough to adopt this approximation: the fraction of energy in the corona was found to be only a few percent even in 70-80\% Pb+Pb collisions \cite{Kanakubo:2019ogh}. An alternative explanation of the same effect is that in small systems, one should use canonical ensemble instead of grand-canonical one \cite{Hamieh:2000tk}. This approach explains the yields of $K$, $\Lambda$, $\Xi$, $\Omega$ in small systems, but fails to explain the yield of $\phi$, which has no open strangeness \cite{Acharya:2018orn}. In our simulation, canonical effects could be implemented using a recently suggested microcanonical sampler \cite{Oliinychenko:2019zfk,Oliinychenko:2020cmr}, but again, in this study, the collision system is large enough to allow us to resort to a usual grand-canonical sampler.

One can see in Fig. \ref{fig:yields} that resonance yields, such as $\rho^0$, $K^{*0}(892)$, $\Lambda(1520)$, are suppressed by the afterburner. In agreement with the experiment, the suppression is more significant in central collisions -- this is consistent with our earlier explanation that the Knudsen number (ratio of resonance mean free path over the system size) plays a role here. A larger system size means a smaller Knudsen number, therefore more scatterings per particle and larger suppression. Resonances with a larger mean free path, $\Sigma(1380)$, $\Xi(1520)$, and $\varphi$ are not suppressed. Their mean free path is large enough to escape the fireball without interactions. It is important to note here that it is not the vacuum lifetime of the resonance that matters, but the in-medium mean free path. This effect can be seen for $\Lambda(1520)$, which has a vacuum lifetime of around 13 fm/$c$, but a small mean free path around 1-2 fm/$c$ in the hadronic medium due to $\pi \Lambda(1520) \leftrightarrow \Sigma^*$ reactions. On the other hand, an alternative Partial Chemical Equilibrium (PCE) model \cite{Motornenko:2019jha}, which does not involve any mean free path considerations, also explains these resonance yields.

The afterburner effect is pronounced in the mean transverse momentum $\langle p_T\rangle$ of the resonances, which one can observe in Fig.~\ref{fig:mpt}. For all shown resonances except $\varphi$, which escapes fireball almost without interactions, hadronic afterburner substantially enhances $\langle p_T\rangle$. This effect can be understood as follows: resonances with small $p_T$ need more time to escape the fireball, and their decay products, which also tend to have small $p_T$, have more time to rescatter. In addition, many cross-sections can be larger for lower relative momenta. Overall this means that the mean free paths of resonances and their decay products decrease for smaller $p_T$. When a resonance or its decay products scatter, we consider it undetectable. Therefore, hadronic scatterings eliminate more resonances with smaller $p_T$, which results in increasing the $\langle p_T \rangle$ of detectable resonances. In addition, the ``pion wind'' effect is acting on resonances accelerating them from low to high $p_T$.

\begin{figure}
    \centering
    \includegraphics[width=0.49\textwidth]{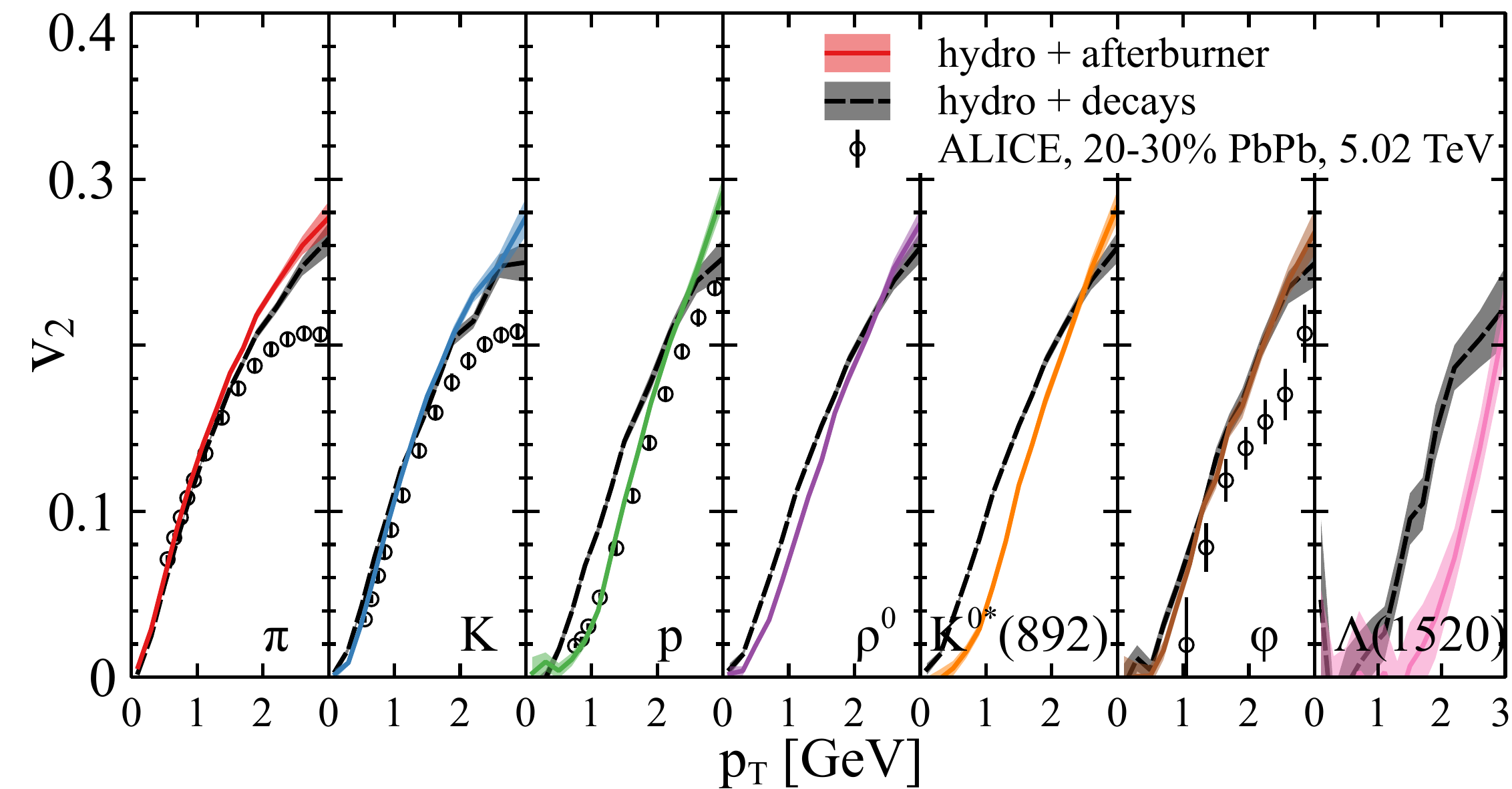}
    \caption{Demonstration of the afterburner effect on the elliptic flow $v_2 = \langle \cos 2\phi_p \rangle $ of hadrons in 20-30\% central Pb+Pb collisions at $\sqrt{s_{\mathrm{NN}}} =5.02$~TeV. Hydrodynamics with afterburner (solid lines) is compared to hydrodynamics with resonance decays (dashed lines) and available experimental data from ALICE \cite{Acharya:2018zuq} (open circles).}
    \label{fig:v2}
\end{figure}

As demonstrated in Fig.~\ref{fig:v2}, the elliptic flow $v_2 = \langle \cos 2 \phi_p \rangle$ of the resonances is also affected by the afterburner. In our simulations, the event plane angle for elliptic flow is zero, $\Psi_2 = 0$. For $\rho^0$, $K^{*0}(892)$, $\Lambda(1520)$, which have small in-medium mean free path, the $v_2$ at low $p_T$ is suppressed: rescattering makes particle's azimuthal distribution more isotropic. For $\varphi$, which has a larger mean free path, the afterburner does not change $v_2$.

\subsection{Extracting hadronic stage duration and \texorpdfstring{$\Sigma^* \to \Lambda(1520)\pi$}{Lg} branching ratios}

\begin{table}[]
    \centering
    \begin{tabular}{lrrrrr}
                   & SMASH      &           & THERMUS &           &  PDG   \\
                   & default    & test 1    & test 2  & test 3    &         \\
$\Sigma(1660) $    &    0.2     &   0       & 0       & 0         &  -      \\ 
$\Sigma(1670) $    &    0.14    &   0       & 0       & 0         &  $>0$     \\
$\Sigma(1750) $    &    0       &   0       & 0       & 0         &  $>0$     \\
$\Sigma(1775) $    &    0.26    &   0.26    & 0.2     & 0         &  0.17-0.23 \\
$\Sigma(1915) $    &    0.59    &   0.59    & 0       & 0         &  -      \\
$\Sigma(1940) $    &    0.17    &   0.17    & 0       & 0         &  $>0$     \\
$\Sigma(2030) $    &    0.195   &   0.195   & 0.15    & 0         &  0.1-0.2 \\                  
    \end{tabular}
    \caption{Branching ratios of $\Sigma^* \to \Lambda(1520)\pi$}
    \label{tab:sigma}
\end{table}

\begin{figure}
    \centering
    \includegraphics[width=0.5\textwidth]{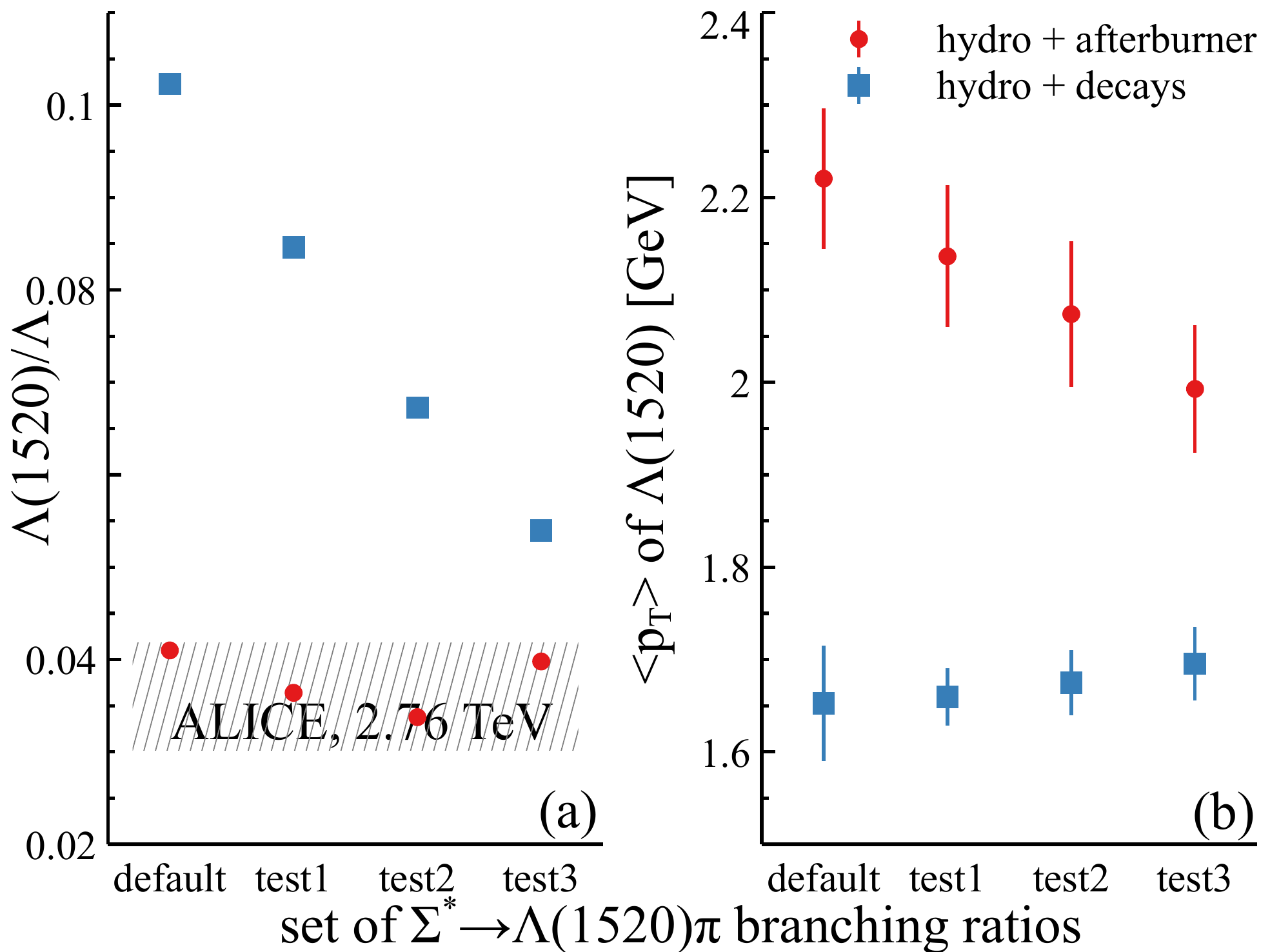}
    \caption{Demonstration of the role of $\Sigma^* \to \Lambda(1520)\pi$ branching ratios in $\Lambda(1520)$ production in 0-10\% central Pb+Pb collisions at 5.02 TeV at midrapidity, $|y|< 0.5$. The ratio $\Lambda(1520)/\Lambda$ (panel a) and the mean transverse momentum of $\Lambda(1520)$ (panel b) in case of hydro + decays (squares) and hydro + afterburner (circles) are compared for different sets of branching ratios $\Sigma^* \to \Lambda(1520)\pi$. These sets are listed in Table~\ref{tab:sigma}, each next test is switching off more branching ratios, in test3 all branching ratios including the ones known experimentally are off. }
    \label{fig:sigmastar_br}
\end{figure}

Out of all considered resonances $\Lambda(1520)$ presents the largest interest, because the influence of afterburner is the most evident, both on its yield, $\langle p_T\rangle$, and elliptic flow (see Figs.~\ref{fig:yields},\ref{fig:mpt},\ref{fig:v2}). The suppression of the yield is consistent with available data. The enhancement of $\langle p_T\rangle$ and the suppression of $v_2$ at small $p_T$ is our prediction. What can we learn from these effects? We suggest two possible answers: one can estimate the duration of the hadronic stage, and one can constrain the branching ratios of $\Sigma^* \to \Lambda(1520)\pi$ decays. The current data seem to require the long lifetime of the hadronic stage (at least 20 fm/$c$) but do not substantially restrict branching ratios $\Sigma^* \to \Lambda(1520)\pi$ decays. The preliminary ALICE data at 5.02 TeV should be sufficient to constrain the branching ratios. Let us elaborate, starting with the branching ratios.

In \texttt{SMASH} there are seven (not counting isospin states) $\Sigma^*$ resonances that decay or may decay into $\Lambda(1520)$. They are listed in Table \ref{tab:sigma} together with the same branching ratios from the Particle Data Group (PDG) summary of the known particle data \cite{Tanabashi:2018oca}. From the PDG column of this table, one can see that only for $\Sigma(1775)$ and $\Sigma(2030)$ these branching ratios are experimentally known. For the rest of $\Sigma^*$ resonances, it is only known that they decay into $\Lambda(1520)$, but the branching ratios are not constrained. The values used in \texttt{SMASH} were set while fitting various strangeness production cross-sections in $pp$, $p\pi$, $pK$, $nK$ collisions \cite{Steinberg:2018jvv}. Systematic uncertainties of the $\Sigma^* \to \Lambda(1520)\pi$ branching ratios were not estimated in this fit. They may be comparable to the branching ratios themselves. To demonstrate how these branching ratios influence $\Lambda(1520)$ midrapidity yield and $\langle p_T \rangle$, we test several sets of branching ratios listed in Table \ref{tab:sigma}. The results are shown in Fig.~\ref{fig:sigmastar_br}. In hydro + resonance decays, larger branching ratios lead to more $\Lambda(1520)$, which is a trivial result. However, it shows by how much a typical thermal model calculation (e.g. \texttt{THERMUS} 3.0 statistical model code \cite{Wheaton:2004qb}) may underestimate $\Lambda(1520)$ yield. Even with this underestimation, statistical models overpredict $\Lambda(1520)$ midrapidity yield because they do not consider the late-stage hadronic rescattering, which strongly suppresses $\Lambda(1520)$ yield. From Fig. \ref{fig:sigmastar_br} it is clear that ALICE 2.76 TeV measurement of the $\Lambda(1520)$ yields \cite{ALICE:2018ewo} is not able to rule out any set of branching ratios in Table~\ref{tab:sigma}. An ongoing analysis of 5.02 TeV data by ALICE is also unlikely to provide more constraints on the $\Sigma^*$ branching ratios from $\Lambda(1520)$ yields. There is, however, a more promising way to constrain the branching ratios -- to consider the $\langle p_T \rangle$ of $\Lambda(1520)$. One can see in Fig.~\ref{fig:sigmastar_br}(b) that $\langle p_T \rangle$ remains almost independent on the branching ratios in the case of hydro + decays but is rather sensitive in the case of hydro + afterburner. Measurements of $\langle p_T \rangle$ with a 5\% precision will be able to distinguish the sets of branching ratios in Table~\ref{tab:sigma}. Meanwhile, as long as the branching ratios are unknown, Fig.~\ref{fig:sigmastar_br} serves as an estimate of the systematic error of our calculation.

\begin{figure}
    \centering
    \includegraphics[width=0.45\textwidth]{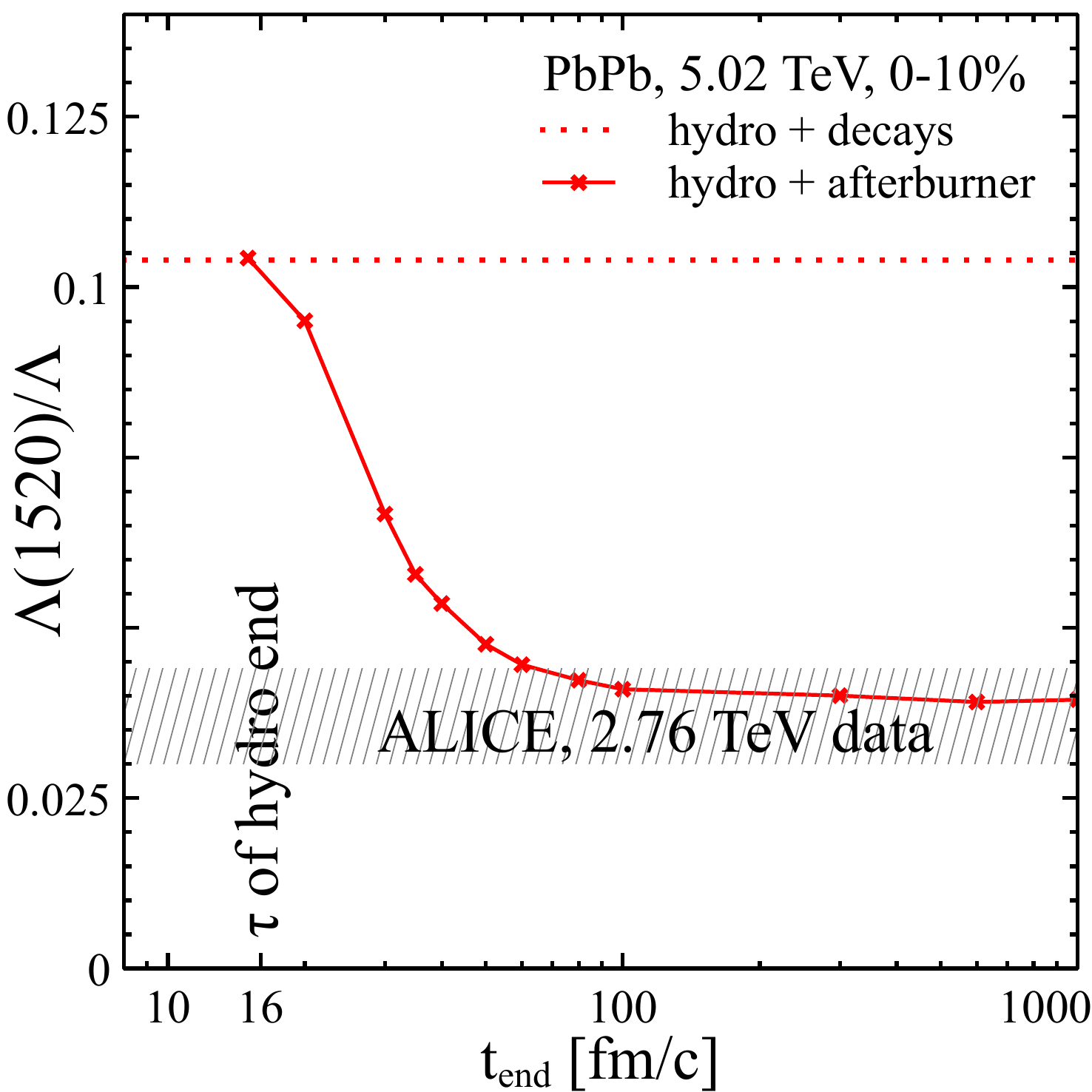}
    \caption{Estimating the duration of hadronic phase from $\Lambda(1520)$ production by stopping simulation at $t_{end}$, and looking at $\Lambda(1520)/\Lambda$ ratio at midrapidity.}
    \label{fig:La1520tend}
\end{figure}

We have already mentioned that the high suppression of $\Lambda(1520)$ allows estimating the duration of the hadronic rescattering stage. Of course, such duration is not rigorously defined in general. One can say that the hadronic stage lasts from chemical to kinetic freeze-out, but both of the freeze-outs are smeared in time and depend on hadron species. Therefore we suggest a model-dependent definition suitable for our model: the duration of the hadronic rescattering stage will be defined as a time interval from the end of the hydrodynamic stage to the moment when all potentially measurable resonance yields do not change by more than 10\% anymore. As $\Lambda(1520)$ yield is suppressed the most by the afterburner among the so far measured resonances, it seems a good proxy for such definition. Therefore, we artificially stop our afterburner simulation early, decay the resonances, and check the $\Lambda(1520)$ yield. As one can see in Fig. \ref{fig:La1520tend}, the earlier one stops the afterburner, the more of $\Lambda(1520)$ remains. The available data are compatible with $t_\mathrm{end} > 40$ fm/$c$, which means that the hadronic stage lasts at least for 20 fm/$c$.

Which reactions contribute to $\Lambda(1520)$ yield decreasing with time? Counting of reactions involving $\Lambda(1520)$ in 0-10\% collisions shows that both the scattering with pions $\Sigma^* \leftrightarrow \Lambda(1520)\pi$, as well as decays and regeneration $\Lambda(1520) \leftrightarrow KN, \Sigma\pi, \Lambda \sigma$ are not equilibrated -- both types of reactions destroy more $\Lambda(1520)$ than create. The rate of $\Sigma^* \leftrightarrow \Lambda(1520)\pi$ is around 10 times larger than $\Lambda(1520) \leftrightarrow KN \,, \Sigma\pi \,, \Lambda \sigma$, but the relative imbalance of the first is much smaller. As a result, both types of reactions contribute to $\Lambda(1520)$ suppression approximately equally. This reaction counting above is for all $\Lambda(1520)$ appearing during the simulation, no matter detectable or not. It is interesting to check from which reactions the detectable $\Lambda(1520)$ originate. It turns out that in 0-10\% collisions for afterburner simulation with default branching ratios around 80\% of final detectable $\Lambda(1520)$ come from a secondary (not born from hydro) $\Sigma^*$ decay; around 15\% come from  $KN \,, \Sigma\pi \,, \Lambda \sigma$ regeneration; and around 5\% come directly from $\Lambda(1520)$ or $\Sigma^*$ sampled at particlization. In peripheral 70-80\% collisions the fraction from $\Lambda(1520)$ or $\Sigma^*$ sampled at particlization increases to around 25\%.

\subsection{Effect of resonance spectral function at sampling}

All the results in Figs. \ref{fig:yields}-\ref{fig:La1520tend} are obtained with sampling resonances at their pole masses. Presently this is a commonly adopted approach, which is justified because the yields of stable hadrons ($\pi$, $K$, $p$, $\Lambda$, $\Xi$, $\Omega$) are not very sensitive to the resonance spectral functions. Protons are affected the most, and their midrapidity yield changes by at most 15\% when spectral functions are included. Depending on the chosen form of the spectral function, the yields can both increase and decrease. The issue has been studied in the thermal model \cite{Bugaev:2013jza,Vovchenko:2018fmh,Andronic:2018qqt} and in the blast-wave model \cite{Huovinen:2016xxq}. Ultimately, the correct way to include spectral functions is by using the experimentally known scattering phase shifts \cite{Andronic:2018qqt}. However, phase shifts are known only for a few reactions. To include spectral functions of all resonances, we take them from our afterburner -- the \texttt{SMASH} transport code. This approach also provides consistency between the sampler and the afterburner. The spectral functions in \texttt{SMASH} are described in detail in \cite{Weil:2016zrk}. They have a relativistic Breit-Wigner shape
\begin{align}
    \mathcal{A}(m^2)dm^2 = \mathcal{N} \frac{2m^2 \Gamma(m)}{(m^2 - m_0^2)^2 + m^2 \Gamma^2(m)} dm \,,
\end{align}
where $m_0$ is the pole mass of the resonance, $\mathcal{N}$ is a normalization factor that provides $\int \mathcal{A}(m^2)dm^2 = 1$, and $\Gamma(m)$ is a mass-dependent width. The total width is composed of the partial widths to all possible decay channels, $\Gamma(m)=\sum_i \Gamma_i(m)$. The mass dependence of the partial width $\Gamma_i(m)$ is rather involved -- it depends on the angular momentum of the decay, and integrates over possible masses of unstable decay products. For more details we refer the reader to \texttt{SMASH} \cite{Weil:2016zrk} and \texttt{GiBUU} \cite{Buss:2011mx} descriptions (\texttt{SMASH} inherits the ideas of resonance treatment from the \texttt{GiBUU} transport code  \cite{Buss:2011mx}).

\begin{figure*}
    \centering
    \includegraphics[width=0.8\textwidth]{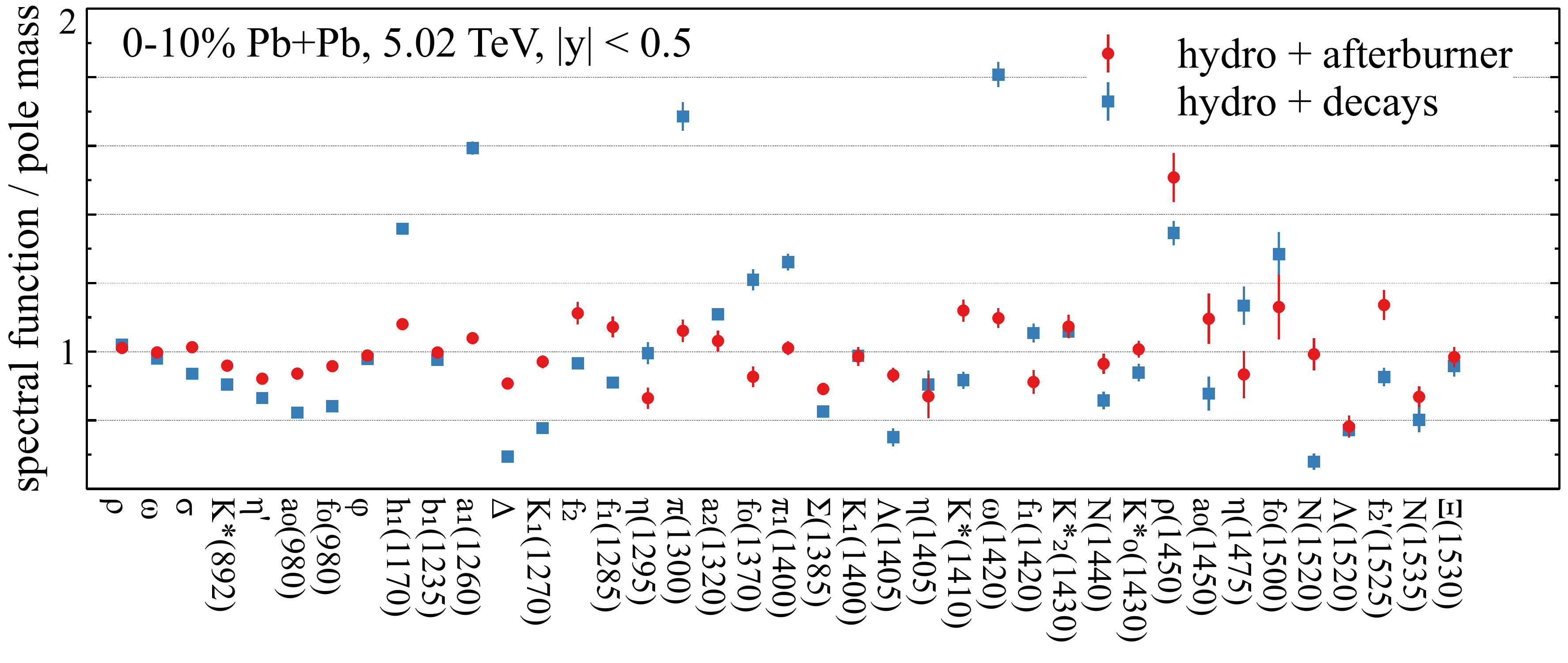}
    \caption{Demonstration of the spectral functions' effects at sampling. Resonance yield ratios are shown -- the numerator is the yield when spectral functions are taken into account in the sampler according to the Eq. \ref{Eq:2}, the denominator is the yield when resonances are sampled at the pole mass according to the Eq. \ref{Eq:1}. The ratio is shown for hydro + decays (squares) and hydro + afterburner (circles). Heavier resonances are not shown to save space.}
    \label{fig:spectral}
\end{figure*}

While stable hadron yields are not affected much by the inclusion of resonance spectral functions, the resonance yields themselves may be, and for some resonances, we find that they are affected substantially. Luckily, the results of the previous sections are not changed by more than 10\% for all resonances considered above, except $\Lambda(1520)$ -- its yield reduces by $\approx 20$ \% when spectral functions are included. Qualitatively all our previous conclusions remain valid. Figure~\ref{fig:spectral} shows that some resonance yields at the sampling are affected by almost a factor of 2 when spectral functions are taken into account. However, after the rescatterings, the spectral functions turn out to be less important -- a typical size of the effect does not exceed 20\%.

The consequence of these results is that any model computing resonance production in heavy-ion collisions and assuming statistical equilibrium -- be it a thermal model, a blast wave model, a hydrodynamic model, or a hydro + afterburner model -- should take resonance spectral functions into account. One can refer to Fig.~\ref{fig:spectral} to estimate the corresponding systematic error, which is typically between 10\% and 30\%, but can be even as large as a factor of 2 for certain resonances.

\subsection{Resonance suppression by afterburner: systematic analysis}

\begin{figure}
    \centering
    \includegraphics[width=0.49\textwidth]{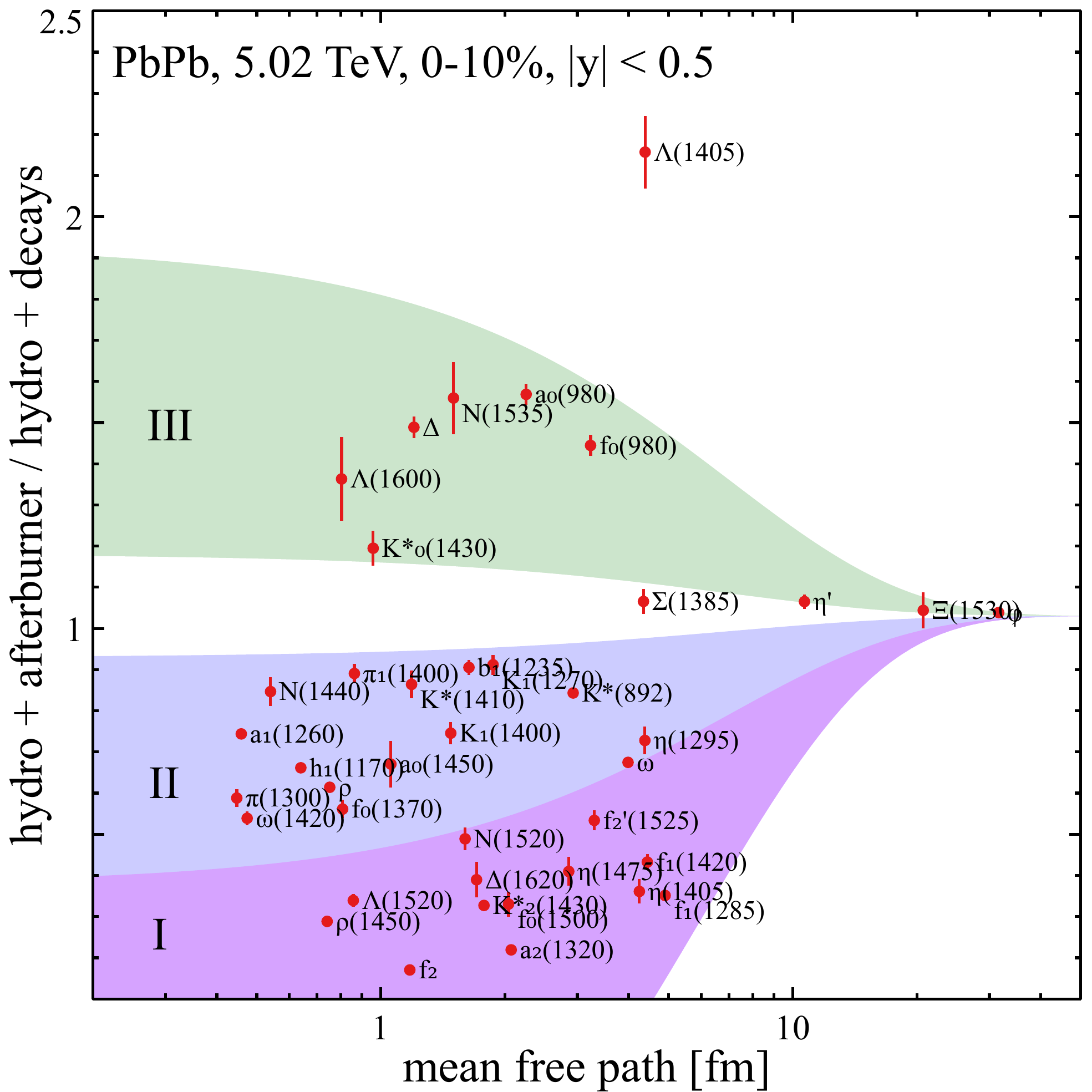}
    \caption{Can the mean free path alone predict suppression of a resonance by the afterburner? This figure demonstrates that it cannot -- the suppression in our simulations is shown against resonance mean free paths computed in a chemically equilibrated medium at temperature $T = 145$ MeV. For explanations of regions I, II, III see text. High mass resonances not shown for better visibility.}
    \label{fig:mfp}
\end{figure}

In previous sections we focused mainly on the resonances that are already measured experimentally: $\rho^0$, $K^{*0}(892)$, $\varphi(1020)$, $\Sigma(1385)$, $\Xi(1530)$, $\Lambda(1520)$. The reasons that out of hundreds of resonances, only these are measured are: (i) they have a decay channel into two measurable charged particles and (ii) the branching ratio of this decay channel is known, (iii) they are relatively narrow, (iv) they do not overlap too much with other resonances. The condition (ii) alone limits a set of potentially measurable resonances to a dozen at most. However, we can trace the full collision history in our simulations, and therefore we are not limited by these conditions. We count any resonance as ``measurable'' if its final decay products reached the end time 100 fm/$c$, while none of the particles along the full decay chain scattered, either elastically or inelastically.

Previously, we qualitatively explained the dependence of resonance suppression on the collision system size by the Knudsen number $Kn$ -- the ratio of a resonance mean free path to the system size. This idea was also helpful to explain why $\varphi$ (long mean free path) is not suppressed, while $\Lambda(1520)$ (short mean free path) is suppressed strongly by hadronic rescattering. Unfortunately, this explanation is not complete. Even at $Kn > 1$, where it seems the most reliable, a resonance can be regenerated from its decay products and enhanced by the afterburner. One can in fact observe this effect in Fig. \ref{fig:yields}(b) for $\varphi$, $\Sigma(1385)$, and $\Xi(1530)$, although the enhancement is small. Already at $Kn \approx 1$, not only the Knudsen number itself matters, the scattering of the resonance decay products also plays a role.
At $Kn \ll 1$ one should expect multiple rescatterings and regenerations, and it is appropriate to use a statistical equilibrium model of expansion, called Partial Chemical Equilibration (PCE) model, where entropy is conserved and stable hadron yields (accounting for stable hadrons ``hidden'' in resonances) are conserved.

\begin{figure}
    \centering
    \includegraphics[width=0.49\textwidth]{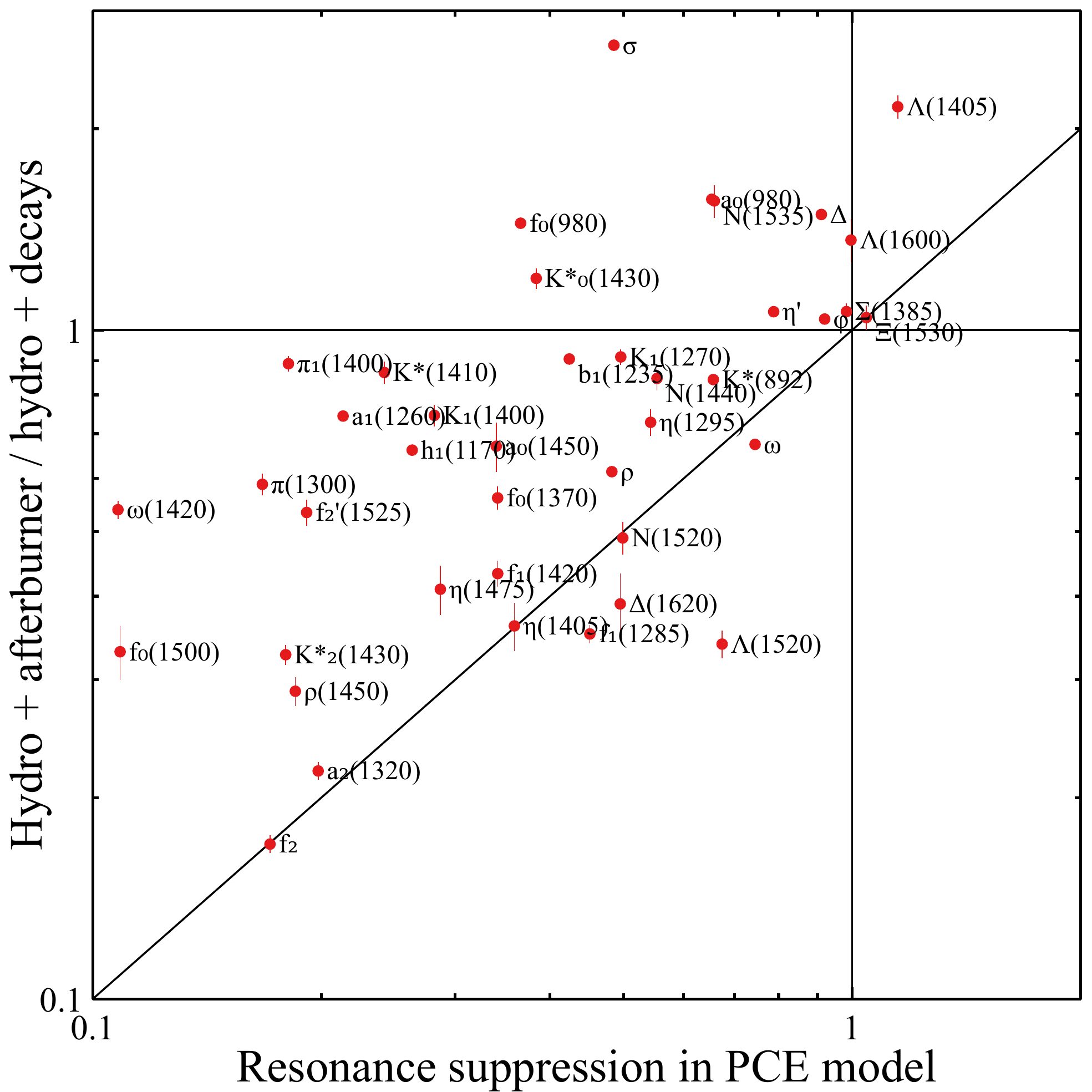}
    \caption{Resonance suppression by the afterburner is compared to the resonance suppression in Partial Chemical Equilibrium (PCE) model. High mass resonances not shown for better visibility.}
    \label{fig:PCE}
\end{figure}

What variable would be a good predictor of a resonance suppression or enhancement by the hadronic stage? We proposed above that the mean free path could be such a predictor. To test our conjecture, for every resonance $R$ we analytically compute mean free path $\lambda_{mfp}$ at $T = 145$ MeV as follows:

\begin{align}
    {\lambda^R_{mfp}}^{-1} = \frac{\langle \Gamma \rangle_{th}}{\hbar c} + \sum_i n^{th}_i \langle \sigma_{iR} v_{rel} \rangle_{th} \,,
\end{align}
where the index $i$ goes over all possible hadrons including resonances and the resonance $R$ itself, $n^{th}_i$ is the density of $i$-th hadron
\begin{align}
    n^{th}_i = \frac{g_i}{(2\pi \hbar c)^3} \int d^3p \, e^{-E/T} \, \mathcal{A}(m^2) dm^2 \,.
\end{align}
The thermally averaged width $\langle \Gamma \rangle_{th}$ takes into account Lorentz time dilation (although here it is not important, because resonances are all non-relativistic at 145 MeV). The $\langle \sigma_{iR} v_{rel} \rangle_{th}$ is thermally averaged inelastic cross section of resonance $R$ with hadron $i$:
\begin{align}
    \langle \Gamma \rangle_{th} = \frac{\int \Gamma(m) \frac{m}{E} d^3p \, e^{-E/T} \mathcal{A}(m^2)dm^2}{\int d^3p \, e^{-E/T} \mathcal{A}(m^2)dm^2} \\
    \langle \sigma_{iR} v_{rel} \rangle_{th} = \frac{\int d^3p_1 d^3p_2 \sigma_{iR}  v_{rel} e^{-(E_1 + E_2)/T} }{ \int d^3p_1 d^3p_2 e^{-(E_1 + E_2)/T}}.
\end{align}

From Fig.~\ref{fig:mfp} it is clear that the mean free path alone does not allow to predict resonance suppression by the afterburner. Indeed, for example $a_2(1320)$ and $a_0(980)$ have similar mean free paths, but $a_2(1320)$ is strongly suppressed, while $a_0(980)$ is substantially enhanced. The mean free path of the resonance does not consider the scattering of the decay products, which is important. Consider two hypothetical resonances $R_1$ and $R_2$ with the same mean free path, but decay products of $R_1$ have very long mean free paths, and decay products of $R_2$ have short mean free paths. In this scenario, $R_1$ will not be suppressed at all, while $R_2$ may be strongly suppressed (or not -- this is not clear a priori). If the decay products rescatter, then one more important factor is if they are likely to regenerate their mother resonance. Unfortunately, we do not find a simple variable to characterize mean free paths of decay products and the tendency to regenerate. However, we make certain qualitative observations using Fig.~å\ref{fig:mfp}:
\begin{itemize}
    \item The less suppressed resonances (denoted as group II in Fig. \ref{fig:mfp}) tend to have a single decay channel or one strongly dominant decay channel, while the more suppressed resonances (group I) tend to have multiple decay channels.
    \item The less suppressed resonances (II) seem to have a larger tendency to regenerate than those in the group (I). For example, $K^*(892)$ decays only into $K\pi$ and when these decay products meet, at temperatures below 145 MeV they more likely create  $K^{*}(892)$ than any other resonances. Similar is valid for $\rho \xrightarrow{100\%} \pi\pi$, $b_1(1235) \xrightarrow{90\%} \omega\pi$, $\pi_1(1400) \xrightarrow{100\%} \pi\eta$, $f_0(1370) \xrightarrow{70\%}\rho\rho$, $\eta(1295) \xrightarrow{100\%}\pi a_0(980)$, which are all in group II. In case of the $\pi\rho$ regeneration all of $a_1(1260),h_1(1170),\pi(1300),\omega(1420)\xrightarrow{100\%} \pi\rho$ are all in the less suppressed group II, probably because of the large pion abundance. This empirical observation has exceptions: for example $\Delta \xrightarrow{100\%} N\pi$ would place $\Delta$ into group II according to our empirical rule, but $\Delta$ is in fact enhanced rather than suppressed.
    \item Enhanced resonances (group III in Fig. \ref{fig:mfp}) tend to be intermediate products of higher mass resonance decays: $\Delta$, $a_0(980)$, $f_0(980)$, $K^*(1430)$, $\sigma$. However, $\Lambda(1520)$ is also an intermediate product, specifically in $\Sigma^*\to\Lambda(1520)\pi$, but it is strongly suppressed and resides in group I.
\end{itemize}

None of these empirical rules is general enough to be satisfactory -- so far, we cannot predict the effect of the afterburner on a resonance yield by some simple analytical calculation or empirical rule without running the afterburner itself. There is, however, one more idea that we would like to test.

Let us consider a limiting case, where mean free paths of all particles are much smaller than the system size. Then frequent collisions keep resonance yields in relative equilibrium with stable hadrons. Also, assume that the yields of stable hadrons (including contributions from resonances) are conserved. In this case, a PCE model introduced in \cite{Motornenko:2019jha} is applicable. Each of the stable hadrons has a corresponding chemical potential $\mu_i$, and resonances have chemical potentials 
\begin{align}
   \tilde{\mu}_j = \sum_{i\in \mathrm{stable}} \langle n_i \rangle_j \mu_i \,,
\end{align}
where $\langle n_i \rangle_j$ is the mean number of a stable hadron $i$ after full decay of a resonance $j$. The chemical potentials $\mu_i(V)$ and temperature $T(V)$ are unknown functions of volume determined from entropy and stable hadron number conservation:
\begin{align}
    \sum_j s_j(T, \tilde{\mu}_j) V = S(T_{ch}) \\
    \sum_j \langle n_i \rangle_j n_j(T, \tilde{\mu}_j) V = N_i(T_{ch}) \,,
\end{align}
where the summation index $j$ runs over all hadrons. We use $T_{ch} = 145$ MeV and stop the fireball expansion at $V/V_{ch} = 3$, which corresponds to a realistic kinetic freeze-out temperature around 96 MeV.

Let us compare the PCE model to our afterburner simulations. Such comparison makes sense for several reasons. Firstly, \texttt{SMASH} strictly fulfills the detailed balance. There exists a reverse reaction for any resonance decay, and matrix elements of any decay and corresponding formation are identical. To provide reserve reactions for $1\to 3$ and $1\to 4$ body decays, they are substituted by a chain of reversible $1\to 2$ decays. Secondly, for many resonances, Knudsen numbers are much smaller than 1. One can conclude it from the mean free paths shown in Fig.~\ref{fig:mfp}. These two conditions provide a possibility that a fireball in our simulation expands in partial chemical equilibrium, which is the assumption of the PCE model. Of course, sooner or later, the fireball becomes too large to sustain the equilibrium, and in the PCE model this complex process is substituted by an assumption of a rapid kinetic freeze-out. Despite this, we would expect that PCE should describe the afterburner results rather well. 

The results of the PCE model are compared to the afterburner simulations in Fig. \ref{fig:PCE}. There is a correlation between them, but it is not sufficient to predict the afterburner effect with at least 20\% accuracy using the PCE model. On average, the PCE model predicts more suppression than the afterburner. One can argue that this is because the chosen kinetic freeze-out volume in PCE is too large. However, in the case of the smaller volume, all resonances in PCE (except $\Xi(1530)$,  $\Lambda(1405)$, and $\Lambda(1600)$) are still suppressed. In contrast, in the afterburner, several resonances are enhanced. We have checked that this enhancement is not related to $2\to 2$ inelastic scattering -- when scatterings are off, and only resonance formation and decays are allowed -- Figure \ref{fig:PCE} does not change substantially in general, and the resonances that were enhanced remain enhanced.

As the $\Delta$ resonance is potentially measurable experimentally, its enhancement by around 50\%, shown in Figs. \ref{fig:mfp} and \ref{fig:PCE}, is particularly interesting. In an earlier hydro + afterburner simulation \cite{Knospe:2015nva} an enhancement of $\Delta$ was observed, but it constituted at most 5\%. The value of 50\% is for the case we account for resonance spectral functions at particlization. In the case of particlization at pole masses, the $\Delta$ enhancement is less significant -- it constitutes around 15\%. One can indeed see in Fig. \ref{fig:spectral} that for $\Delta$ the spectral function effect is only around 0.9 for hydro + afterburner, and around 0.7 for hydro + decays, and $\frac{0.9}{0.7} \approx \frac{1 + 50\%}{1 + 15\%}$. To ensure that our result does not originate from an unlikely detailed balance violation, we use the test particles method with $N_{test} = 10$: oversample by a factor of 10 at particlization and reduce all scattering cross sections by factor 10. Such a procedure is known to reduce detailed balance problems (if there are any) substantially. Generally, the larger $N_{test}$, the closer results of the simulation should approach the solution of the corresponding Boltzmann equation. After introducing $N_{test} = 10$ we found that the changes in resonance production, including $\Delta$, are within 5\%. This result gives us confidence that enhancement of $\Delta$ is not a result of some unexpected detailed balance violation in \texttt{SMASH} code. Altogether we predict an enhancement of $\Delta$ resonance in central collisions.

\section{Summary} \label{sec:Summary}

We have studied resonance production in Pb+Pb collisions at 5.02 TeV using the hydrodynamics + hadronic afterburner simulation. The simulation reproduces the measured midrapidity yield ratios $\rho^0/\pi$, $K^{0*}/K^-$, $\Sigma(1385)/\Lambda$, $\Lambda(1520)/\Lambda$, $\Xi(1530)/\Xi$, $\varphi/K$ reasonably well as a function of collision centrality. We make predictions for the mean transverse momentum and the $p_T$-differential flow $v_2(p_T)$ of the resonances. We confirm that the suppression of $\rho$, $K^*$, and $\Lambda(1520)$ in central collisions results from a late-stage hadronic rescattering, which is presently a rather well-established conclusion. The enhancement of $\langle p_T\rangle$ of resonances by the hadronic afterburner, which we observe in our simulations, is also a known effect. The suppression of resonance $v_2$ at small $p_T$ is a new result from our simulations.

As $\Lambda(1520)$ production is affected particularly strongly by the hadronic rescattering stage, we explored what one can learn from measuring it. The measurement of $\Lambda(1520)/\Lambda$ midrapidity yield ratio allows estimating the duration of the hadronic stage. The measurement of $\langle p_T\rangle$ of $\Lambda(1520)$ will help to constrain the $\Sigma^* \to \Lambda(1520) \pi$ branching ratios.

Unlike previous theoretical works, we have analyzed not only the production of experimentally accessible resonances, but also the production of all resonances included in the simulation. We focused on the question: ``can one predict the afterburner effect on resonance production in a simpler way than running the full afterburner simulation?'' Resonance vacuum lifetime and regeneration cross-section are known to be poor predictors of resonance suppression. Resonances' mean free paths turn out to have predictive power only when the mean free path is large. We noticed some empirical rules that tend to be fulfilled with certain exceptions. In particular, resonances that have one dominant decay channel tend to be suppressed less, presumably because of the regeneration. Some resonances ($a_0(980)$, $f_0(980)$, $\Delta$, $N(1535)$, $\Lambda(1405)$, $N(1535)$, $K^*_0(1430)$) are enhanced by afterburner -- this is our prediction, and it will be interesting to test experimentally. Interestingly, all enhanced resonances are the intermediate products of higher resonance decays; but it is not true vice versa. The Partial Chemical Equilibrium model \cite{Motornenko:2019jha}, which agrees with afterburner simulation for the measured resonances, cannot predict resonance suppression or enhancement if a larger set of resonances is considered. Altogether, we have not found a simple predictor of a resonance suppression (or enhancement) by the hadronic rescattering stage. In the absence of a better approach the PCE model remains the least inaccurate approximation to the full simulation of hadronic rescattering.

\acknowledgements
The authors thank L. McLerran, V. Koch, A. Sorensen, S. Pratt, and V. Vovchenko for useful comments.  C.~S. was supported in part by the U.S. Department of Energy (DOE)  under grant number DE-SC0013460 and in part by the National Science Foundation (NSF) under grant number PHY-2012922. D.O. was supported by the U.S. DOE under Grant No. DE-FG02-00ER4113.
This work is supported in part by the U.S. Department of Energy, Office of Science, Office of Nuclear Physics, within the framework of the Beam Energy Scan Theory (BEST) Topical Collaboration.
Computational resources were provided by the high performance computing services at Wayne State University, and by Goethe-HLR computing cluster.

\bibliography{inspire,noninspire}
\end{document}